\documentclass[doublecol]{epl2}

\usepackage{amssymb,amsmath,latexsym}

\newcommand\beq{\begin{equation}}
\newcommand\eeq{\end{equation}}

\newcommand\bea{\begin{eqnarray}}
\newcommand\eea{\end{eqnarray}}

\newcommand\bi{\begin{itemize}}
\newcommand\ei{\end{itemize}}

\newcommand\non{\nonumber}

\newcommand\up{\uparrow}

\newcommand\odd{1{\textendash}D}

\newcommand\od{1{\textendash}D~}

\newcommand\ie{{\it{i.e.}}}

\newcommand\smat{$\mathbb S$}

\newcommand\ctd{{\textsf{CT}}}

\newcommand\ct{{\textsf{CT~}}}

\newcommand\db{{\textsf{DB~}}}

\newcommand\scb{{\textsf{SB~}}}

\newcommand\sdbd{{\textsf{SDB}}}

\newcommand\sdb{{\textsf{SDB~}}}

\newcommand\nsd{{\textsf{NS}}}

\newcommand\nsnd{{\textsf{NSN}}}

\newcommand\scurrentd{{\textsf{SC}}}

\newcommand\card{{\textsf{CAR}}}
\newcommand\ard{{\textsf{AR}}}

\newcommand\ed{{\textsf{e}}}
\newcommand\hd{{\textsf{h}}}

\newcommand\ns{{\textsf{NS~}}}

\newcommand\nsn{{\textsf{NSN~}}}

\newcommand\scurrent{{\textsf{SC~}}}

\newcommand\car{{\textsf{CAR~}}}
\newcommand\ar{{\textsf{AR~}}}

\newcommand\bdgd{{\textsf{BdG}}}
\newcommand\bdg{{\textsf{BdG~}}}

\newcommand\e{{\textsf{e~}}}

\def\dfrac#1#2{{\displaystyle\frac{#1}{#2}}}

\newif\ifboo \boofalse
\def\Review#1{\boofalse{\it #1},}

\def\Vol#1{\ifboo Vol. {\bf #1}\else{\bf #1}\fi}
\def\Year#1{\ifboo #1\else(#1)\fi}

\def\Page#1{\ifboo {\rm p. #1}\else{\rm #1}\fi}

\title{Resonant spin transport through a superconducting double barrier
structure}
\shorttitle{Resonant spin transport through a \sdb structure}

\author{Arijit Kundu \and Sumathi Rao \and Arijit Saha}
\shortauthor{Arijit Kundu \etal}

\institute{Harish$-$Chandra  Research Institute,
Chhatnag Road, Jhusi, Allahabad 211 019, India}

\abstract{We study resonant transport through a superconducting double
barrier structure. At each barrier, due to the proximity effect, an
incident electron can either reflect as an electron or a hole
(Andreev reflection). Similarly, transport across the barrier can
occur via direct tunneling as electrons as well as via the crossed Andreev
channel, where a hole is transmitted. In the subgap regime, for a
symmetric double barrier system (with low transparency
for each barrier), we find a new $T=1/4$ resonance
($T$ is the transmission probability for electrons incident on the double
barrier structure) due to  interference between electron and hole
wave-functions between the two barriers, in contrast to a normal
double barrier system which has the standard transmission resonance at
$T=1$. We also point out as an application that the resonant value of $T=1/4$
can produce pure spin current through the superconducting double barrier
structure.}

\begin{document}

\maketitle

\section{Introduction}
Effects due to the proximity of a superconductor have motivated a
lot of research in the recent past both from
theoretical~\cite{blonder,lambert,beenaker,beenakkerns,fazioreview,morpurgo,
hekking1,
belzig2006,zaikin12007,das2007drsahaprb} as well as
experimental~\cite{russo,chandrasekhar} point of view. Due to the proximity
effect, an electron incident on a normal metal-superconductor (\nsd)
interface reflects back as a hole and as a consequence, two electrons are
transferred into the superconductor as a Cooper pair. This phenomenon is
known as Andreev reflection (\ard)~\cite{andreev} in the literature of
mesoscopic superconductivity. An even more intriguing example where the
proximity effect manifests itself is the phenomenon of crossed Andreev
reflection (\card)~\cite{hekking1,belzig2006,zaikin2007,zaikin12007} in
which an electron from one of the normal leads of a normal
metal-superconductor-normal metal (\nsnd) junction pairs up with another
electron from the other lead and as a result, a Cooper pair jumps into the
superconductor. This nonlocal process can only take place if the
separation between the points of coupling of the two normal metal leads
with the superconductor is of the order of the size of the Cooper pair
itself. From the application point of view, the relevance of \car in the
manipulation of pure spin currents (\scurrentd)~\cite{das2007drsahaepl},
spin filter~\cite{Chtchelkatchev} and production of entangled electron
pairs in nanodevices~\cite{recher,thierrymartin1,yeyati} has attacted a
lot of interest in recent times.

Motivated by these facts, in this letter, we adopt the simple-minded
definition of \scurrent which is commonly used in
literature~\cite{review,rashba,psharma}. It is just the product of the
local spin polarization density associated with the electron or hole, (a
scalar $s$ which is either positive for spin-up or negative for spin-down)
and its velocity. To generate a pure \scurrent in the sense defined above,
one can have the two most obvious scenarios where $(a)$~there
exists an equal and opposite flow of oppositely spin-polarized electrons
through a channel, such that the net charge current through the channel is
nullified leaving behind a pure \scurrent, or $(b)$~alternatively, there
exists an equal flow of identically spin polarized electrons and holes in
the same direction through a channel giving rise to pure \scurrent with
perfect cancellation of the charge current. In this letter, we explore the
second possibility for generating resonant pure \scurrent using a
superconducting double barrier (\sdbd) structure.


\begin{figure*}[htb]
\begin{center}
\includegraphics[width=10cm,height=6cm]{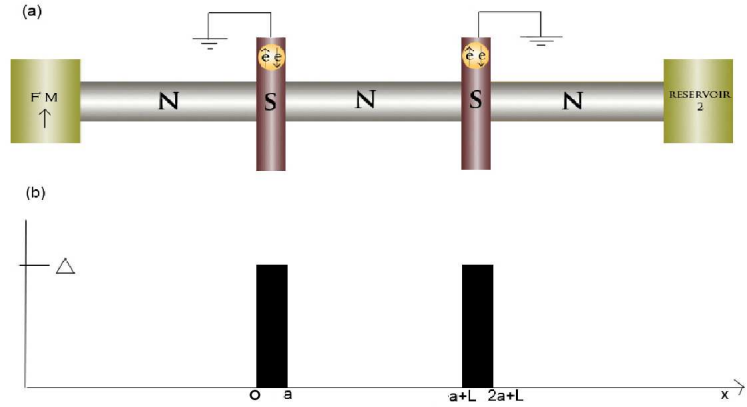}
\caption{(a)~Cartoon of the \sdb structure where a single channel,
ballistic, \od normal metal (N) lead is connected to a
ferromagnetic (F) reservoir at its left end and to a normal reservoir
at its right. Two patches at the two places on the lead depict
superconducting material deposited on top of it. (b)~Schematic of the
potential profile seen by an incident electron.}
\label{geometry}
\end{center}
\end{figure*}

\section{Proposed device and its theoretical modelling}
The configuration we have in mind for the generation of resonant pure
\scurrent is shown in Fig.\ref{geometry}. The idea is to design a \sdb
structure by depositing thin strips of superconducting material on top of
a single channel, ballistic, one-dimensional (\odd) lead at two places,
which can induce a finite superconducting gap ($\Delta_{i}e^{i \phi_{i}}$)
in the barrier regions as a result of proximity effect of the
superconducting strips. If the width of the strips is of the order of the
phase coherence length of the superconductor, then both direct electron to
electron co-tunneling (\ctd) as well as crossed-Andreev electron to hole
tunneling can occur across the \sdb region. Here it is worth mentioning
that we restrict ourselves to  spin singlet superconductors so that
both elastic \ct and \car across the junction conserve spin. In our
theoretical modeling of the system (see Fig.~\ref{geometry}),
we first assume that the \smat-matrix representing the \sdb structure
described above respects parity (left-right symmetry) and 
spin-rotation symmetry, so that we can describe the system by
an \smat-matrix with only eight independent parameters namely, {\it{(i)}}
the normal refection amplitude ($r_c$ ($\tilde{r}_c$)) for \e (\hd), 
{\it{(ii)}} the transmission or \ct amplitude ($t_c$ ($\tilde{t}_c$)) for 
\e (\hd), {\it{(iii)}} the Andreev reflection (\ard) amplitude ($r_{Ac}$ 
($\tilde{r}_{Ac}$)) for \e (\hd), and {\it{(iv)}} the
crossed Andreev reflection (\card) amplitude ($t_{Ac}$ ($\tilde{t}_{Ac}$)) 
for \e (\hd). (We have used the subscript $c$ to denote the composite 
amplitudes for the \sdb structure and tilde to distinguish the amplitudes 
for incident holes from the incident electron).
If we inject spin polarized electron ($\up$ \ed) from the left lead using a
ferromagnetic reservoir and tune the system parameters such that $t_c$ and
$t_{Ac}$ are equal to each other, it will lead to a pure \scurrent flowing
to the right lead. This is so because, on an average, an equal number
of electrons ($\up$ \ed) (direct electron to electron \ctd) and holes
($\up$ \hd) (\car of electron to hole tunnelling) are injected from the
left lead to the right lead resulting in the cancellation of the average
charge current, whereas the spins add, giving rise to pure \scurrent in
the right lead. Note that spin up holes ($\up$ \hd) implies a Fermi sea
with an absence of spin down electron (which is needed for the incident
electron $(\up \ed)$ to form a Cooper pair and jump into the singlet
superconductor).

\begin{figure*}[htb]
\begin{center}
\includegraphics[width=10cm,height=8cm]{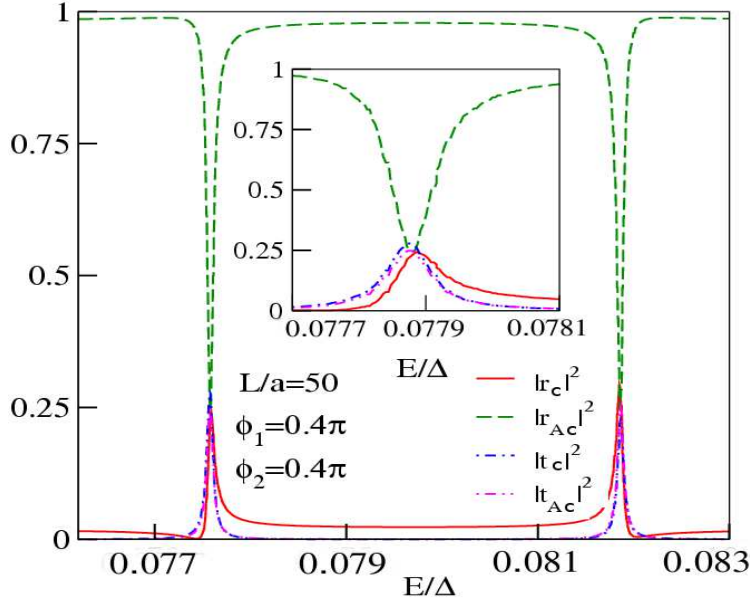}
\caption{The variation of $|r_c|^{2},|t_c|^{2},|r_{Ac}|^{2}$ and $|t_{Ac}|^{2}$
is plotted in units of $e^{2}/h$ as a function of $E/\Delta$ for a
symmetric \sdb system. $\Delta /E_F$ has been chosen to be 0.275.
Inset:~The region close to one of the resonances is expanded in the inset.
Note the existence of two closely spaced resonances.}
\label{sdbres}
\end{center}
\end{figure*}

\section{Superconducting double barrier}
Quantum transport in the \sdb structure has been studied earlier by Morpurgo
\etal~in Ref.~\cite{morpurgo}. In their work, they assumed that both the
barriers were reflectionless and also that there was no \car across
either of the barriers. They then obtained the resultant Andreev
reflection and transmission amplitudes across the \sdb by considering
multiple \ar processes between the barriers and found a $T=1$ resonance.
In this letter, we address the full problem allowing all the
quantum mechanical processes occuring across the two barriers. Hence
our set-up is very similar to that given in Ref.~\cite{morpurgo},
and comprises of a ballistic normal \od lead with two short, but
finite superconducting patches deposited on top of it as shown in
Fig.\ref{geometry}. Here the structure is connected to ideal
ferromagnetic and normal electron reservoirs respectively at its two ends.
$\Delta_{i}$ and $\phi_{i}$ are the pair potentials and order parameter
phases on the two patches respectively ($i$ refers to the index of the
strips). The space dependence of the order parameter (which also acts as a
scattering potential) for the incident electron can be expressed as

\begin{eqnarray} \Delta(x) &=& \Delta^{(i)} e^{i\phi_1} \Theta(x)
\Theta(-x+a) + \Delta^{(i)}e^{i\phi_2} \non\\
&&  \Theta[x-(a+L)] \Theta[-x+(2a+L)]
\end{eqnarray}
where, $a$ is the width of the \scb, $L$ is the distance between the
two barriers and $\Theta$ is the Heaviside $\Theta$ function.

In contrast to Ref.\cite{morpurgo}, to obtain the resultant reflection,
transmission,
\ar and \car for the \sdb structure, one
needs to consider all the multiple reflection processes in the
\sdb due to both $r_i$ and $r_{Ai}$ for each of the barriers labelled by $i$.
So an electron which enters the region between the two barriers has a
choice of being reflected as an electron or being converted to a hole at
each bounce. For the numerical analysis, it is more convenient
to use the
alternate method to solve such scattering problems, which is to use the
standard wave-function matching technique.

The one-dimensional Bogolubov$-$de Gennes (\bdgd) equation~\cite{degennes}
for (spin up and spin down) electrons and holes can be written as
\bea Eu_{+} &=&
\left[\dfrac {-\hbar^{2} \nabla^{2}} {2m} + V(x) - \mu_{L} \right]u_{+}
+ \Delta u_{-} \eea \bea Eu_{-} &=& \left[\dfrac {\hbar^{2}
\nabla^{2}} {2m} - V(x) + \mu_{R} \right]u_{-} + \Delta^\star u_{+}
\label{bdgeqn}
\eea
The solution of the \bdg equations, describing electrons and holes with
incident energy $E$ inside the normal regions ($\Delta_{(i)}=0$), can be
written as
\bea
\Psi_{e}^{\pm q^{+}x}(x)&=&\Bigg(\begin{array}{c} 1 \\ 0 \end{array} \Bigg)
{e}^{\pm iq^{+}x}
\label{enormal}
\eea
\bea
\Psi_{h}^{\mp q^{-}x}(x)&=&\Bigg(\begin{array}{c} 0 \\ 1 \end{array} \Bigg)
{e}^{\mp iq^{-}x}
\label{hnormal}
\eea
where, $\hbar q^{\pm}=\sqrt{2m(E_{F}\pm E)}$ and the $\pm$ sign in the
exponent of the plane wave solutions corresponds to an excitation
propagating in the $\pm x$ direction.

Similarly, inside the superconducting barrier regions the solutions for
electronlike and holelike excitations are
\bea
\Psi_{e}^{\pm k_i^{+}x}(x)&=&\Bigg(\begin{array}{c} u_{i+}e^{\pm i \phi_{i}}
\\ u_{i-} \end{array} \Bigg)
{e}^{\pm ik_i^{+}x}
\label{esupcon}
\eea
\bea
\Psi_{h}^{\mp k_i^{-}x}(x)&=&\Bigg(\begin{array}{c} u_{i-}e^{\pm i \phi_{i}}
\\ u_{i+} \end{array} \Bigg){e}^{\mp ik_i^{-}x}
\label{hsupcon}
\eea
where, $\hbar k_i^\pm = \sqrt{2m(E_F \pm (E^2-\Delta_{(i)}^{2})^{1/2})}$,
$u_{\pm}=\frac {1}{\sqrt{2}}[(1\pm(1-(\Delta_{(i)}/E)^{2})^{1/2})]^{1/2}$,
$m$ is the effective mass of the electron and $E_{F}$ is the Fermi energy
of the system.

Hence matching the wavefunctions for the normal and superconducting
regions (Eq.(\ref{enormal}-\ref{hsupcon})) at the four \ns interfaces
($x=0,a,a+L,2a+L$) forming the \sdb structure, we obtain sixteen linear
equations. Numerically solving these sixteen equations we obtain the
4$\times$4 \smat-matrix for the \sdb structure which, for an incident
electron with energy $E$, can be written as
\bea \mathbb{S}_{e} =
\begin{bmatrix} ~r_c & t_c& r_{Ac} & t_{Ac}~\\
~t_c & r_c & t_{Ac} & r_{Ac}~\\
~r_{Ac} & t_{Ac} & r_c & t_c~\\
~t_{Ac} & r_{Ac} & t_c & r_c~\\
\end{bmatrix}
\label{smatsdb}
\eea
In Eq.\ref{smatsdb} $r_c$ stands for normal reflection of electrons or holes
and $r_{Ac}$ represents \ar (reflection of an electron as a hole or
vice-versa) from the barriers. Similarly, $t_c$ represents \ct or normal
transmission amplitude of electrons or holes while $t_{Ac}$ represents the
nonlocal \car amplitude for electron to hole conversion across the \sdb
structure.The amplitudes depend on the incident energy $E$, the Fermi
energy $E_F$ and the length $L$ between the barriers.

For the normal double barrier system, resonant electron transport occurs
whenever $\theta =\pm 2q^+ L = n\pi$, which is the condition for
quasi-bound states inside the double barrier. The situation for the \sdb
system is much more subtle. Since both electrons and holes are bounced
in the normal region between the two barriers, there are multiple
path dependent phases. For instance, an electron that gets reflected
as an electron
gets a phase of $2q^+ L$, a hole that gets reflected as a hole gets a
phase of $2q^- L$,
and an electron that gets Andreev reflected as a hole gets not only the
path dependent phase of $(q^+-q^-)L$, but also a $\Delta$ dependent phase of
$\cos^{-1}E/\Delta$~\cite{beenaker}.

We now obtain the resonance condition by using the technique
of adding up all the Feynman paths that contribute to the transmission
amplitude. However, here since we have both transmission and \car, we need
to use matrices for reflection and transmission.
Let us assume
the reflection and transmission matrices at each of the
two superconducting barriers to
be the
same and given by
\beq
{R} = \begin{pmatrix}
r_e&r_{Ah}\chi\\
r_{Ae}\chi & r_h
\end{pmatrix}  \quad {\rm and} \quad  {T} = \begin{pmatrix}
t_e&t_{Ah} \\
t_{Ae} & t_h
\end{pmatrix}
\eeq
Note that we have allowed for  the amplitudes to be
electron -electron and hole-hole reflections and transmissions
$r_{e(h)}$ and $t_{e(h)}$ to be different  and also similarly
electron-hole ($r_{Ae}$ and $t_{Ae}$) amplitudes to
be different from hole-electron ($r_{Ah}$ and $t_{Ah}$) amplitudes.
The $\Delta$-dependent phase $\chi = e^{-i\cos^{-1}E/\Delta}$~\cite{beenaker}
has also been included along with each Andreev reflection.
The path-dependent phases can also be conveniently written in a
matrix form as
\beq
P = \begin{pmatrix}
\eta & 0\\
0 & \nu
\end{pmatrix}
\eeq
where
$\eta = e^{iq^+L}$ and $\nu = e^{-iq^- L}$
are  the phases picked up by the electron and hole respectively,
as they  move a distance $L$.

In terms of these matrices, if $I_{R,L}$  and
$O_{R,L}$ (each of them are column
vectors denoting electrons and holes) are the
incoming and outgoing  waves moving towards the left
or right we find
\beq
\begin{pmatrix}
O_L \\
O_R \end{pmatrix} = \begin{pmatrix} {R} + {TP}R Q {PT}&
{T}Q {PT} \\
{T}Q {PT} & R+ TPRQ{PT} \end{pmatrix}
 \begin{pmatrix} I_R \\ I_L \end{pmatrix}
\label{comp}
\eeq
where $Q =(I- PRPR)^{-1}$. We define $D = {\rm Det} (I-(PR)^2)$.
The condition for resonant transport can now be easily found
from the composite transmission amplitude. For an incident electron
going from left to right
with wave-function given by $I_R= (1,0)$ ($I_L =(0,0)$),
the amplitude for an electron to be transmitted towards the right
(upper component of $O_R$) is given by the 1-1 component
of the matrix $TQPT = TPQT$. The explicit expression for
the amplitudes are cumbersome to display, particularly
since the electron - hole symmetry is broken.
Similarly, the composite amplitudes for \car (1-2 component of $TQPT$), 
reflection (1-1 component of $R+TPRQPT$) and \ar (1-2 component of $R+TPRQPT$) 
can also be found from the above matrix.

Clearly, the condition for resonant transport is now set by the vanishing
of the denominator - \ie., when $D(E=E_r+iE_i) = 0$,
$|t_c|^2(E_r)$ has a maximum.
Note that the composite amplitudes for all the 4 processes
have the same denominator and hence show resonant behaviour
when the denominator goes to zero. We find that all four of
them (using the correct expressions from the matrix)
become 1/4 at resonance, which is a maximum for $|t_{c}|^2,|t_{Ac}|^2$
and $|r_{c}|^2$ and a minimum for $|r_{Ac}|^2$.
Note also that setting $r_{A(e,h)}=t_{A(e,h)}=0$
give the usual double barrier resonance condition, whereas setting
$r_{(e,h)}=t_{A(e,h)}=0$ gives the resonance
studied by Morpurgo and Beltram~\cite{morpurgo}.

Our model does not explicitly include any external barrier at any
of the normal-superconductor or superconductor-normal interfaces.
In the earlier study of \sdb without any barrier~\cite{morpurgo},
the approximation $\Delta/E_F \ll 1$ was also taken, which led
to the vanishing of $r$, and consequently $t_A$. However,
when $\Delta/E_F$ is less than unity, but not vanishingly
small ($e.g.$, we have taken $\Delta/E_F$ between $1/4 - 1/10$),
back-scattering and hence a small non-zero value for $r_i$ (which
is the reflection at each barrier) does exist.
We have also checked that the inclusion of normal
barriers (two external $\delta$-function impurities at each \ns interface)
along with the superconducting barriers does not change the result
substantially.
\begin{figure*}[htb]
\begin{center}
\vskip +0.5cm
\includegraphics[width=13.5cm,height=6.2cm]{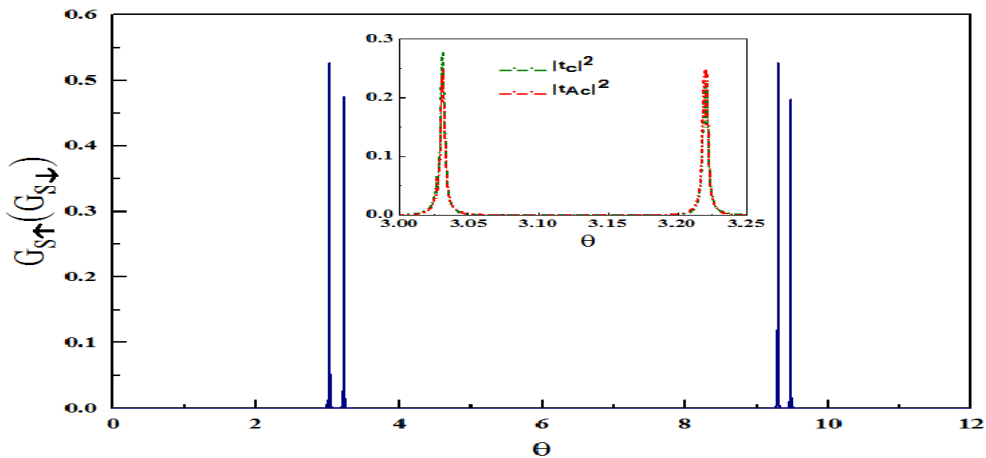}
\caption{The behaviour  of the spin conductance $ \varpropto
|t_c|^2 + |t_{Ac}|^2$ for spin $\uparrow$ ($\downarrow$) polarised
electrons, is shown, in units  of the incident spin,  as a function of
$\theta=(q^{+}-q^{-})L$ for a
symmetric \sdb system. The parameter values are given by  $E/\Delta = 0.078$,
$\Delta/E_F = 0.275$ and $\phi_1=\phi_2 = 0.4\pi$. The inset shows
the variation of $|t_c|^{2}$ and $|t_{Ac}|^{2}$  close to the resonance, in
units of $e^{2}/h$. Note the closely spaced resonances in the inset. }
\label{sdbpsc}
\end{center}
\end{figure*}

\section{Results}
In this section we describe the consequences of all the allowed quantum
mechanical processes across the \sdb given by the \smat-matrix in
Eq.\ref{smatsdb}.

\subsection{Resonance structure:-}
As mentioned earlier, we numerically solve the  16 linear equations
obtained by matching wave-functions at the four \ns junctions. We restrict
ourselves to the subgap regime, where  electron energy $E$ is much less
than the gap energy ($E \ll \Delta_{i}$ and $r_i$ is small).
For a symmetric \sdb system,
$\Delta_{1}=\Delta_{2}=\Delta; \phi_{1}=\phi_{2}$. The behaviour of
$|r_c|^{2}, |t_c|^{2}, |r_{Ac}|^{2}$ and $|t_{Ac}|^{2}$ as a function of
$E/\Delta$ is shown in Fig.~\ref{sdbres}. Note that for some particular
values of $E/\Delta$, the coherent probabilities for all the
\smat-matrix amplitudes given in Eq.\ref{smatsdb} become $1/4$.
Note also that the graphs show two closely spaced resonances.
This can be understood from the analytic expression in Eq.\ref{comp},
and more specifically from the denominator $D$.
For small values of $r$, there are two resonances slightly displaced
from the doubly degenerate
pure Andreev level resonances which occurs when $r=0$.
Furthermore, for  $E \ll E_F$, $q^+ \simeq q^- \simeq q^F = \sqrt
{2 m E_F}$, and the $r$ term in the determinant
for fixed $E_F$ has no significant phase
dependence. However, the $r_A$ term is multipled by the
phase $e^{i(q^+-q^-)L}$.
Hence, we have plotted the variation
of the transmission and \car probabilities as a function of
$\theta = (q^+ - q^-)L$, in Fig.\ref{sdbpsc}(a). Note the approximate
periodicity of the resonances is for $\theta \rightarrow \theta+2\pi$.

The width of the resonance depends on the back-scattering that
occurs at a single barrier, which in turn, depends on the
value of $\Delta/E_F$ (for $r_{Ai}$) and the scattering
potential at the barrier (for $r_{i}$). We have checked
numerically that changing $\Delta/E_F$ between $1/4 - 1/10$
changes $r_{Ai}$ which affects the width of the resonances, and changing
the strength of the $\delta$-function impurities changes $r_i$, but otherwise
has very little affect on the character of the graph.
Hence, we have presented
our results only for $\Delta/E_F = 0.275$ and  $\lambda=0$, where
$\lambda$ is the strength of the $\delta$-function.

Setting $r_i=0$ exactly reverts to the
problem studied by Morpurgo and Beltram~\cite{morpurgo}
which has only a $T=1$ resonance. This is
similar to the usual $|t|^{2}=1$ resonance of a standard normal \db system,
(which is obtained by setting $r_{Ai}=0$)~
although the physical origin is different, since for the \sdb structure,
the electron gets Andreev reflected at each bounce between the barriers,
instead of getting normally reflected. As long as the multiple bounces
between the barriers occur either through normal reflection or Andreev
reflection, but not both, the transmission resonance remains unimodular.
On the other hand in our case, we have allowed for all the quantum
mechanical processes that can occur at each barrier. Hence, we have
multiple bounces between the two barriers involving both $r_i$ and
$r_{Ai}$.  This seems to lead to a new non-unimodular resonance
at $T=1/4$, where in fact, all the quantum mechanical probabilities
become $1/4$ because all the 4 processes show resonant
behaviour. {\it This is the main point of this letter.}
It is hence clear that the very occurence of
the $T=1/4$ resonancee
requires the presence of all the four amplitudes \ie, in a `single'
channel problem with just one reflection and one transmission,
the only resonance that is possible is the standard unimodular
resonance. The non unimodular $T=1/4$ resonance requires the
presence of two `channels'.

\subsection{Pure spin current:-}
As an application of the above \sdb geometry, we point out
that this geometry can be used to produce pure \scurrent in a resonant
fashion. The proposal for generating pure spin current using \nsn junction
was discussed earlier in Ref.\cite{das2007drsahaepl}, but there it
involved non resonant production of pure \scurrent unlike the present case.
In our analysis the {\it{spin conductance}} is defined as
$G_{S\uparrow}(G_{S\downarrow}) \varpropto |t_{Ac}|^{2}+|t_c|^{2}$
in units of the incident spin,
whereas the {\it{charge conductance}} is given by
$G_{C\uparrow}(G_{C\downarrow}) = (e^2/h)[ |t_{Ac}|^{2}-|t_c|^{2}]$.
The $\uparrow$ and $\downarrow$ arrows in the subscript represent
the spin polarization of the injected electrons from the ferromagnetic
reservoir as shown in Fig.\ref{geometry}. The sum of contributions coming
from two oppositely charged particles (electrons and holes) gives rise to
the negative sign in the expression for
$G_{C\uparrow}(G_{C\downarrow})$. The interesting point to note here
is that for an electron incident on the barriers, if the amplitudes for
the \ct and \car are identical, then it will result in equal probability
for an incident electron to transmit as an electron or as a hole across
the barriers. This results in the vanishing of the charge
current. On the other hand, in our geometry, if the incident electron in
the lead is $\uparrow$ or $\downarrow$ spin polarized, then both the
transmitted electron due to $t_c$ and hole due to $t_{Ac}$ will have the same
spin polarization. This is true because in our analysis we have
assumed that the superconducting patches are spin singlets and hence spin
remains conserved. Note, however, that if the incident electrons were
not spin polarised (\ie, if we  had a normal metal
reservoir instead of a ferromagnet), then even when \ct and \car
are equal, there
would be both up and down spin electrons and holes transmitted, and hence
there would be no \scurrent.
Therefore, when the symmetric \sdb structure with a ferromagnetic reservoir
is tuned
to resonance \ie~$|t_c|^{2}=|t_{Ac}|^{2}=\dfrac{1}{4}$ and if a spin polarized
beam (say $\uparrow$ spin polarized according to Fig.\ref{geometry}) of
electrons is incident on the barriers, then the outcome would be
resonant production of outgoing pure \scurrent. In this resonant
situation $25\%$ of the incident spin-up electrons get transmitted through
the barriers via the \ct process and $25\%$ get converted to spin-up
holes via the \car process as they pass through the barriers. Hence the
transmitted charge across the barriers is zero on the average, but there
is pure \scurrent flowing out of the system.
The behaviour of $G_{S\uparrow}(G_{S\downarrow})$ for the \sdb
system as a function of $(q^+ - q^-) L $ is shown in Fig.\ref{sdbpsc}(b).
At the resonance, $G_{S\uparrow}(G_{S\downarrow})$ becomes $0.5$ and
$G_{C\uparrow}(G_{C\downarrow})$ becomes  $0$ for a spin
polarized electron beam which is a clear manifestation of pure
\scurrent in a \sdb geometry.
\section{Discussions and Conclusions}
In this letter, we have studied a superconducting double barrier system
and have shown that one can tune a $T=1/4$ resonance in the system.
It is crucial to have non-zero amplitudes for all the four
amplitudes, reflection, Andreev reflection, normal transmission
and crossed Andreev reflection to see this resonance.
Note also that we have restricted ourselves to
the  Blonder-Tinkham-Klapwijk approximation~\cite{blonder}
of neglecting the single
electron transfer across the barrier. Hence, we have
shown the resonance only in the thick barrier limit, where
the transparency is low.
A similar resonance
was already noted by some of us~\cite{arijitstub} in a stub geometry where
all these four amplitudes were non-zero. The special value
of $T=1/4$ was also
noted earlier by some of us~\cite{das2007drsahaepl,
das2007drsahaprb} in the context of a weak interaction
renormalisation group study of a \nsn junction, where a non-trivial
fixed point was found which had $T=1/4$. In all these contexts,
not only does $T$ have a non-unimodular value, all the other
amplitudes also have a value of 1/4,
($|t_{Ac}|^2 = |r_c|^2 = |r_{Ac}|^2 =1/4$)
as required by unitarity.
In all these contexts, pure \scurrent is the outcome (at resonance or at
the fixed point), since the charge current gets nullified on the average.

As far as the practical realization of such a \sdb structure is concerned,
it should be possible to fabricate such a geometry by depositing thin
strips of a spin singlet superconductor (like $Nb$ with $\Delta \sim$1.5meV~\cite{russo})
on top of a ballistic
quantum wire (with $E_F \sim$1eV~\cite{yacoby1}) or a carbon nanotube at two places. The
width of the strips should be of the order of the superconducting phase
coherence length ($10-15 nm$ in case of $Nb$). The $T=1/4$ resonance in
this \sdb geometry can be tuned by varying the energy of the incident
electron (which can be done by applying a small bias voltage between the
two reservoirs keeping within linear response, so that our calculations
are valid) for fixed distance ($\sim 0.5 \mu m$) between
the two barriers or the distance between the two barriers for fixed
incident energy.
However, inclusion of electron$-$electron interaction, finite temperature,
finite bias, etc can lead to very interesting physics in the presence of
resonances which is beyond the scope of the present work.

In conclusion, in this letter we have studied resonant transport through
a \sdb stucture where, at an energy scale much below the superconducting
gap $\Delta$, probablities for all the coherent amplitudes become $1/4$
in the tunneling (or thick barrier) approximation.
As an application we have also discussed the possibility for production of
resonant pure \scurrent in this structure.
\section{Acknowledgements}
\vspace{.2cm}
We thank Sourin Das for many stimulating and useful discussions and also
a careful reading of the manuscript. We acknowledge use of the Bewoulf
cluster at HRI for our numerical computations.
\bibliographystyle{eplbib} 
%
\bibliography{superdbref} 
\end{document}